\documentclass[prd,aps,nofootinbib,preprint,tightenlines]{revtex4}

\newcommand{\checked}[1]{}

\usepackage{graphicx}

\newcommand{\beq}{\begin{equation}}
\newcommand{\eeq}{\end{equation}}
\newcommand{\bqa}{\begin{eqnarray}}
\newcommand{\eqa}{\end{eqnarray}}
\def\simge{\mathrel{
    \rlap{\raise 0.511ex \hbox{$>$}}{\lower 0.511ex \hbox{$\sim$}}}}
\def\simle{\mathrel{
    \rlap{\raise 0.511ex \hbox{$<$}}{\lower 0.511ex \hbox{$\sim$}}}}

\begin{document}

\title{Flow Effects on Jet quenching with Detailed Balance}


\author{Luan Cheng$^{a,b}$ and Enke Wang$^{a,b}$}
\affiliation{$^a$Institute of Particle Physics, Huazhong Normal
University, Wuhan 430079, China\\
$^b$Key Laboratory of Quark $\&$ Lepton Physics (Huazhong Normal\\
University), Ministry of Education, China}

\begin{abstract}
A new model potential in the presence of collective flow describing
the interaction of the hard jet with scattering centers is derived
based on the static color-screened Yukawa potential. The flow effect
on jet quenching with detailed balance is investigated in pQCD. It
turns out that the collective flow changes the emission current and
the LPM destructive interference comparing to that in the static
medium. Considering the collective flow with velocity $v_z$ along
the jet direction, the energy loss is $(1 - v_z )$ times that in the
static medium to the first order of opacity. The flow dependence of
the energy loss will affect the suppression of  high $p_T$ hadron
spectrum and anisotropy parameter $v_2$ in high-energy heavy-ion
collisions.
\end{abstract}
\maketitle

\small

\section{Introduction}

One of the most striking features of nucleus-nucleus collisions at
the Relativistic Heavy Ion Collider (RHIC) is the collective flow.
In recent years this phenomenon has been a subject of intensive
theoretical and experimental
studies\cite{Lorstad,Voloshin1,Wiedemann1,Voloshin2}. It is believed
that the medium produced in nucleus-nucleus collsions at RHIC
equilibrates efficiently and builds up a flow field.

Gluon radiation induced by multiple scattering of an energetic
parton propagating in a dense medium leads to induced parton energy
loss or jet quenching. As discovered in high-energy heavy-ion
collisions at RHIC, jet quenching is manifested in both the
suppression of single inclusive hadron spectrum at high transverse
momentum $p_T$ region\cite{phenix} and the disappearance of the
typical back-to-back jet structure in dihadron
correlations\cite{star}. Extentive theoretical investigation of jet
quenching has been widely carried out in recent
years\cite{GW94,BDPMS,Zakharov,GLV,Wiedua,GuoW}. Most of the jet
quenching theory research are studied in a static medium based on
the static color-screened Yukawa  potential proposed by Gyulassy and
Wang\cite{GW94}.
However, the medium is not static, the collective flow need to be
considered\cite{BDMS,SW,WW}.
Later, the interaction between the jet and the target partons in the
presence of collective flow was modeled by a momentum shift
$\mathbf{q}_0$ perpendicular to the jet direction in the
Gyulassy-Wang's static potential\cite{ASWiedemann}, but this
assumption lacks sufficient theoretical evidence. Recently, local
transport coefficient $\hat{q}$, which is related to the squared
average transverse momentum transfer from the medium to the hard
parton per unit length, has been investigated in the presence of
transverse flow\cite{BMS,Liu}. However, when relating $\hat{q}$ with
the energy density $\varepsilon$ of the medium, $\hat{q}\simeq
c\varepsilon^{3/4}$, a problem appears that the determination of $c$
is different from $c=2$ to $c>8,\cdots,19$ by various authors, the
results lack consistency. The $\hat{q}$ study with collective flow
in Ref.\cite{BMS} based on BDMPS energy loss calculation gives only
a macroscopic result for parton energy loss. Many interesting
properties in jet quenching theory such as the flow effects on the
non-Abelian Laudau-Pomeranchuk-Migdal(LPM) interference effect and
opacity can not be studied. On the other hand, only radiative energy
loss can be considered in Ref.\cite{BMS}, the detailed balance
effect with gluon absorption cannot be included. It has been shown
that the gluon absorption play an important role for the
intermediate jet energy region\cite{Wang:2001sf}.

In this letter, we report a first study of the parton energy loss
with detailed balance in the presence of collective flow in
perturbative Quantum Chromodynamics (pQCD). We first determine the
model potential to describe the interaction between the energetic
jet and the scattering target partons with collective flow of the
quark-gluon medium using Lorentz boosts. Based on this new
potential, we then consider both the radiation and absorption
induced by the self-quenching and multiple scattering in the moving
medium. We are led to the conclusion that to the zeroth order
opacity, the energy loss is dominated by the final-state thermal
absorption, whose result is the same as that in the static medium
since the jet has no interaction with the medium. However, for the
case of rescattering with targets, the collective flow changes the
emission current and the LPM destructive interference. Overall, it
reduces (enhances) the jet energy loss induced by rescattering with
stimulated emission and thermal absorption depending on the
direction of the flow in the positive (negative) jet direction.

\section{The Potential Model}

To calculate the induced radiation energy loss of jet in a static
medium, the interaction potential is assumed in the Gyulassy-Wang's
static model \cite{GW94} that the quark-gluon medium can be modeled
by N well-separated color screened Yukawa potentials,
\begin{eqnarray}
V_i^a(q_i)=2\pi\delta(q_i^0)\frac{4\pi\alpha_s}
{\mathbf{q}^2+\mu^2}e^{-i\mathbf{q}\cdot \mathbf{x_i}}
T_{a_i}(j)T_{a_i}(i) \label{GWpotential}\, ,
\end{eqnarray}
where $\mu$ is the Debye screening mass, $T_{a_i}(j)$ and
$T_{a_i}(i)$ are the color matrices for the jet and target parton at
$\mathbf{x}_i$. In this potential, each scattering has no energy
transfer ($q_i^0=0$) but only a small momentum $\mathbf{q}$ transfer
with the medium. If using the four-vector potential, the
Gyulassy-Wang's static potential can be denoted as
$A^{\mu}=(V_i^a(q_i),\mathbf{A}(q_i)=0)$.

As is well known in Electrodynamics, the static charge produces a
static Coulomb electric field, while a moving charge produces both
electric and magnetic field.  In analogy a moving target parton in
the quark-gluon medium will produce color-electric and
color-magnetic fields simutaneously due to the collective flow.
Therefore, the static potential model should be reconsidered.

In the quark-gluon medium with collective flow, the rest frame fixed
at target parton moves with a velocity $\mathbf{v}$ relative to the
observer's system frame $\Sigma'$, as illustrated in Fig.1. We first
take a Lorentz transformation for four-momentum $q$, and then for
four-vector potential $A^{\mu}$, we can then write
$A^{\mu}=(V_{i(flow)}^a(q_i),\mathbf{A}_{(flow)}(q_i))$ in the
observer's system frame $\Sigma'$ as
\begin{equation}
\left\{
\begin{array}{ll}
V_{i(flow)}^a(q_i){=}2\pi\delta(q_i^0{-}\mathbf{v}\cdot\mathbf{q})
e^{-i\mathbf{q}\cdot
\mathbf{x}_i} \tilde{v}(\mathbf{q})T_{a_i}(j)T_{a_i}(i)\, , \\
\mathbf{A}_{(flow)}(q_i){=}2\pi\delta(q_i^0{-}\mathbf{v}\cdot\mathbf{q})
\mathbf{v} e^{-i\mathbf{q}\cdot
\mathbf{x}_i}\tilde{v}(\mathbf{q})T_{a_i}(j)T_{a_i}(i)\, ,
\end{array}
\right.
 \label{flowpotential}
\end{equation}
where
$\tilde{v}(\mathbf{q})={4\pi\alpha_s}/({\mathbf{q}^2-(\mathbf{v}\cdot\mathbf{q})^2+\mu^2})$.
The new potential differs from Gyulassy-Wang's static potential in
that the collective flow of the quark-gluon medium produces a
color-magnetic field and the flow leading to non-zero energy
transfer $q_i^0=\mathbf{v}\cdot\mathbf{q}$, which will affect jet
energy loss as we will show below.

\begin{figure}
\begin{center}
\includegraphics[width=63mm]{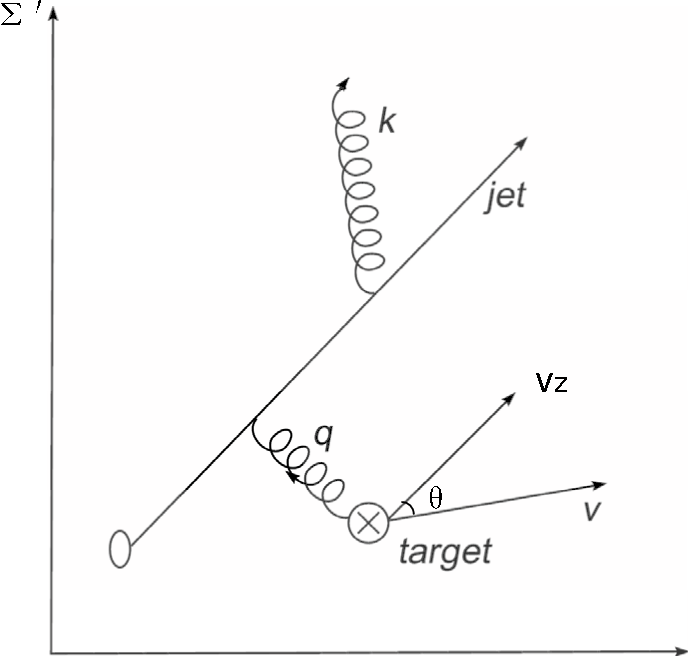}
\end{center}
\vspace{-12pt} \caption{\label{fig:pic1} View of kinematics in the
quark-gluon medium with collective flow. }
\end{figure}

Elastic cross section for small transverse momentum transfer between
jet and target partons can be deduced as
\begin{equation}
\frac{d\sigma_{el}}{d^2 \mathbf{q}}=\frac{C_RC_2}{d_A}\frac{\mid
\tilde{v}(\mathbf{q}) \mid ^2 }{(2\pi)^2}\, , \label{elasticcrosec}
\end{equation}
where $C_R$ and $C_2$ are the Casimir of jet and target parton in
fundamental representation in $d_R$ dimension, respectively. $d_A$
is the dimension of corresponding adjoint representation. Our result
agrees with the GLV elastic cross section in static potential when
the flow velocity goes to zero \cite{GW94}.

\section{Flow Effect on Gluon Radiation}

Consider a hard parton produced at
$\tilde{z}_0=(z_0,\mathbf{x}_{\perp})$. The hard parton has initial
energy E, interacts with the target parton at
$\tilde{z}_1=(z_1,\mathbf{x}_{1\perp})$ with flow velocity
$\mathbf{v}$ by exchanging gluon with four-momentum $q$, radiates a
gluon with four-momentum $k$ and polarization $\epsilon(k)$, and
emerges with final four-momentum $p$. In the light-cone components,
\begin{eqnarray}
k &= &[2\omega,\frac{\mathbf{k}_{\perp}^2}{2\omega},
  \mathbf{k}_{\perp}]\, ,\\
 \epsilon(k) &=& [0,2\frac{\mathbf{\epsilon}_{\perp}\cdot
 \mathbf{k}_{\perp}}{xE^+},\mathbf{\epsilon}_{\perp}]\, ,\\
p &=& [(1-x)E^++2\mathbf{v}\cdot
\mathbf{q},\mathbf{p}_-,\mathbf{p}_{\perp}^+]\, ,
 \label{lightconecom}
\end{eqnarray}
where $\omega=xE$, $E^+=2E\gg\mu$.

At zeroth order in opacity, the jet has no interaction with the
target parton, we obtain the same factorized radiation amplitude off
a quark
 \begin{equation}
 \label{rad0}
   R^{(0)}=2ig T_c
   \frac{{\bf k}_{\perp}\cdot{\mathbf \epsilon}_{\perp}}
   {{\bf k}^2_{\perp}}\, ,
 \end{equation}
as that in the static medium in Ref. \cite{Wang:2001sf}, where $T_c$
is the color matrix. As shown in Ref.\cite{Wang:2001sf}, the net
energy gain without rescattering can be expressed as
 \begin{equation}
 \label{elossab0}
   {\Delta E^{(0)}_{abs}}{\approx} {-}
   \frac{\pi\alpha_s C_R}{3}
   \frac{T^2}{E}\left[
     \ln\frac{4ET}{\mu^2}{+}2{-}\gamma_{\rm E}
     {+}\frac{6\zeta^\prime(2)}{\pi^2}\right],
\end{equation}
where $\gamma_{\rm E}\approx 0.5772$, $\zeta^\prime(2)\approx
-0.9376$ and $T$ is the thermal finite temperature.

However, at first order in opacity, consider the jet has the
simplest case of elastic scattering. The radiation amplitude
 \begin{eqnarray}
 \label{rad1}
   M^{(1)} &\propto& -i(2p-q)_{\mu}A^{\mu} \nonumber \\
           &\propto& 2\pi\delta(q_i^0{-}\mathbf{v}\cdot\mathbf{q})
e^{-i\mathbf{q}\cdot \mathbf{x}_i}
T_{a_i}(j)T_{a_i}(i)(2E+\mathbf{v}\cdot \mathbf{q})R^{(1)}
           \, ,
 \end{eqnarray}
 where $R^{(1)}=(1-v_z)\tilde{v}(\mathbf{q})$, which is changed by
 the collective flow with a factor $(1-v_z)$.

When the hard parton goes through the quark-gluon medium, it will
suffer multiple scattering with the parton target inside the medium.
Here we investigate the rescattering-induced radiation by
considering the flow effect resulting from the moving parton target.
We will work in the framework of opacity expansion developed by
Gyulassy, L\'{e}vai and Vitev (GLV)\cite{GLV} and
Wiedemann\cite{Wiedua}. It was shown by GLV that the higher order
corrections contribute little to the radiative energy loss. So we
will only consider the contributions to the first order in the
opacity expansion. The opacity is defined as the mean number of
collisions in the medium, ${\bar n}\equiv
L/l=N\sigma_{el}/A_{\perp}$. Here $N$, $L$, $A_{\perp}$ and $l$ are
the number, thickness, transverse area of the targets, and the
average mean-free-path for the jet, respectively.

Based on our new potential in Eq.(\ref{flowpotential}) by
considering collective flow of the quark-gluon medium, assuming the
flow velocity $|\mathbf{v}|\ll 1$, we obtain the factorized
radiation amplitude associated with a single rescattering,
\begin{eqnarray}
R^{(S)}&=&2ig\Bigl(\mathbf{H}T_aT_c+\mathbf{B_1}e^{\frac{i\omega_{0}z_{10}}
{1-v_z}}[T_c,T_a] -2v_zT_aT_c
\mathbf{H}(1-e^{\frac{i\omega_{0}z_{10}}{1-v_z}})
+\mathbf{C_1}e^{\frac{i(\omega_{0}-\omega_{1})z_{10}}{1-v_z}}
  [T_c,T_a] \Bigr)\cdot {\mathbf \epsilon}_{\perp}\nonumber \\
  && \times(1-v_z)\, ,
\label{singleamp}
\end{eqnarray}
where $z_{10}=z_1-z_0$,
\begin{eqnarray}
 \label{omega}
   \omega_0&=&\frac{{\bf k}_{\perp}^2}{2\omega}\, ,
   \quad
   \omega_1=\frac{({\bf k}_{\perp}{-}{\bf q}_{\perp})^2}
      {2\omega}\, ,
 \\
   {\bf H}&=&\frac{{\bf k}_{\perp}}{{\bf k}_{\perp}^2}\, ,
   \quad
   {\bf C}_1=\frac{{\bf k}_{\perp}{-}{\bf q}_{\perp}}
      {({{\bf k}_{\perp}{-}{\bf q}_{\perp}})^2}\, ,
   \quad
   {\bf B}_1={\bf H}{-}{\bf C}_1\, .
 \end{eqnarray}
Different from the static medium case, the single rescattering
amplitude depends on the collective flow of the quark-gluon medium.

The interference between the process of double scattering and no
rescattering should also be taken into account to the first order in
opacity \cite{GLV}. Assuming no color correlation between different
targets, the double rescattering corresponds to the ``contact
limit'' of double Born scattering with the same target \cite{GLV}.
Assuming the flow velocity $|\mathbf{v}|\ll 1$, with our new
potential in Eq.(\ref{flowpotential}) the radiation amplitude can be
expressed as
\begin{eqnarray}
 \label{ampd}
   R^{(D)}&=&2ig T_c e^{\frac{i\omega_0z_{10}}{1-v_z}}
   \Bigl(-{{C_R+C_A}\over 2}{\bf H} e^{-\frac{i\omega_0z_{10}}
   {1-v_z}}
    +{C_A\over 2}{\bf B}_1
    +2v_z\frac{C_R-C_A}{2}{\bf H}(1-e^{-\frac{i\omega_0z_{10}}{1-v_z}})
 \nonumber \\
    &&
     +{C_A\over 2}{\bf C}_1 e^{-\frac{i\omega_1z_{10}}{1-v_z}}\Bigr)
     \cdot {\mathbf \epsilon}_{\perp}(1-v_z)^2,
\end{eqnarray}
where $C_A$ is the Casimir of the target parton in adjoint
representation in $d_A$ dimension. The double rescattering amplitude
also depends on the collective flow.

To the first order in opacity, we then derive the induced radiation
probability including both the stimulated emission and thermal
absorption as
\begin{eqnarray}
 \label{probafirstord}
   {{dP^{(1)}}\over d\omega}&=&{{C_2}\over {8\pi d_A d_R}}{N\over A_{\perp}}
     \int {{dx}\over x}  \int
      {{d^2{\bf k}_{\perp}}\over {(2\pi)^2}}
     \int {{d^2{\bf q}_{\perp}}\over {(2\pi)^2}} P({\omega\over E}) |R^{(1)}|^2\left\langle
      Tr\left[|R^{(S)}|^2{+}2
      Re\left(R^{(0)\dagger}R^{(D)}\right)\right]\right\rangle
     \nonumber\\
     &&
     \Big[\left(1{+}N_g(xE)\right)\delta(\omega{-}xE)\theta(1{-}x)
      {+}N_g(xE)\delta(\omega{+}xE)\Big]
      \nonumber \\
   &\approx & {{\alpha_s C_2 C_R C_A}\over {d_A \pi}}{N\over A_{\perp}}
     \int {{dx}\over x}  \int
      {{d{\bf k}_{\perp}^2}\over {\bf k}_{\perp}^2}
     \int {{d^2{\bf q}_{\perp}}\over {(2\pi)^2}} P({\omega\over
     E})J^2_{eff}(k_{\perp},{\bf q}_{\perp})
       \left\langle Re(1{-}e^{\frac{i\omega_1z_{10}}{1{-}v_z}})\right\rangle
      \nonumber \\
        &&
         \Big[\left(1{+}N_g(xE)\right)\delta(\omega{-}xE)\theta(1{-}x)
         +N_g(xE)\delta(\omega+xE)\Big]\, ,
\end{eqnarray}
where $N_g(|{\bf k}|)=1/[\exp(|{\bf k}|/T)-1]$ is the thermal gluon
distribution, $v({\bf q}_{\perp})={{4\pi\alpha_s}/ ({{\bf
q}^2_{\perp}+\mu^2}})$, $\alpha_{s}=g^2/4\pi$ is strong coupling
constant. We have also included the splitting function
$P_{gq}(x)\equiv P(x)/x=[1+(1-x)^2]/x$ for $q\rightarrow gq$.

The gluon formation factor $1-\exp(i\omega_1z_{10}/(1-v_z))$
reflects the destructive interference of the non-Abelian LPM effect.
The formation time of gluon radiation $\tau_f\equiv
(1-v_z)/\omega_1$ becomes shorter (longer), the LPM effect is
reduced (enhanced) in the presence of collective flow in the
positive (negative) jet direction, respectively. The gluon formation
factor must be averaged over the longitudinal target profile, which
is defined as $\left\langle \cdots\right\rangle=\int dz
\rho(z)\cdots$. We take the target distribution as an exponential
Gaussian form $\rho(z)=\exp(-z/L_e)/L_e$ with $L_e=L/2$, the gluon
formation factor can be deduced as
\begin{eqnarray}
 \label{LPMfactor}
 \left\langle Re(1{-}e^{\frac{i\omega_1z_{_{10}}}{1{-}v_z}})\right\rangle
 &=&
\frac{2}{L}\int_0^{\infty}d  z_{_{10}} e^{-{2z_{_{10}}}/{L}}
(1{-}e^{{\frac{i\omega_1z_{_{10}}}{1{-}v_z}}}) \nonumber \\
&=&\frac{(\mathbf{k_{\perp}}-\mathbf{q_{\perp}})^4L^2}{16x^2E^2(1-v_z)^2
+(\mathbf{k_{\perp}}-\mathbf{q_{\perp}})^4L^2}\, .
\end{eqnarray}

Here the emission current $J_{eff}(k_{\perp},{\bf q}_{\perp})$ in
Eq.(\ref{probafirstord}) is defined as
\begin{eqnarray}
 \label{emissioncurrent}
J^2_{eff}(k_{\perp},{\bf q}_{\perp})
       &=&{|R^{(1)}|^2\left\langle
      Tr\left[|R^{(S)}|^2{+}2
      Re\left(R^{(0)\dagger}R^{(D)}\right)\right]\right\rangle} \over {\left\langle
      Re(1{-}e^{\frac{i\omega_1z_{10}}{1{-}v_z}})\right\rangle}
      \nonumber \\
      & \approx & (1-v_z)^2{{{\bf k}_{\perp}\cdot {\bf q}_{\perp}} \over
      {\left({\bf k}_{\perp} - {\bf q}_{\perp}\right)^2}}\, .
\end{eqnarray}
It shows that the collective flow reduces the emission current with
a factor $(1-v_z)$ when $0<v_z<<1$.

The jet energy loss can be divided into two parts. The
zero-temperature part corresponds to the radiation induced by
rescattering without detailed balance effect and can be expressed as
\begin{eqnarray}
 \label{energyloss1}
\Delta E^{(1)}_{rad}&=&\int d\omega \omega {{dP^{(1)}}\over
d\omega}\Big|_{T=0}
\nonumber \\
&=&\frac{\alpha_s C_R}{\pi}\frac{L}{l_g} E \int dx \int {{d{\bf
k}_{\perp}^2}\over {\bf k}_{\perp}^2} \int {d^2{\bf q}_{\perp}}
|{\bar v}({\bf q}_{\perp})|^2 P(x)J^2_{eff}(k_{\perp},{\bf
q}_{\perp})
      \left\langle
      Re(1{-}e^{\frac{i\omega_1z_{10}}{1{-}v_z}})\right\rangle,
\end{eqnarray}
where $l_g=C_R l/C_A$ is the mean-free path of the gluon.

The temperature-dependent part of energy loss induced by
rescattering at the first order of opacity comes from thermal
absorption with partial cancellation by stimulated emission, in the
presence of flow it can be written as
\begin{eqnarray}
 \label{eabs1}
   \Delta E^{(1)}_{abs}&=&\int d\omega \,\omega
     \left({{dP^{(1)}}\over d\omega} -{{dP^{(1)}}\over d\omega}
       \Big|_{T=0}\right) \nonumber \\
       &=&\frac{\alpha_s C_R}{\pi}\frac{L}{l_g} E \int dx \int {{d{\bf
k}_{\perp}^2}\over {\bf k}_{\perp}^2} \int {d^2{\bf q}_{\perp}}
|{\bar v}({\bf q}_{\perp})|^2 N_g(xE) J^2_{eff}(k_{\perp},{\bf
q}_{\perp})
\nonumber \\
&&
      \big[ P({-}x)\left\langle
      Re(1{-}e^{\frac{i\omega_1z_{10}}{1{-}v_z}})\right\rangle
   -P(x)\left\langle
      Re(1-e^{\frac{i\omega_1z_{10}}{1-v_z}})\right\rangle\theta(1-x)\big],
 \end{eqnarray}
where $|{\bar v}({\bf q}_{\perp})|^2$ is defined as the normalized
distribution of momentum transfer from the scattering centers,
 \begin{eqnarray}
 \label{vbar}
  &&|{\bar v}({\bf q}_{\perp})|^2 \equiv {1\over \sigma_{el}}
  {d^2\sigma_{el}\over d^2{\bf q}_{\perp}}=
  {1\over \pi}{\mu^2_{eff}\over ({\bf q}_{\perp}^2+\mu^2)^2}\, ,
 \\
   &&{1\over \mu^2_{eff}} = {1\over \mu^2}-{1\over
   q_{\perp max}^2+\mu^2}\, ,\quad q_{\perp max}^2\approx 3E\mu\, .
 \end{eqnarray}


To obtain a simple analytic result, we take the kinematic boundaries
limit $q_{\perp max}\rightarrow\infty$, the angular integral can be
carried out by partial integration. In the limit of $EL \gg 1$ and
$E\gg \mu$,  we obtain the approximate asymptotic behavior of the
energy loss,
 \begin{equation}
 \label{elossem3}
   {\Delta E_{rad}^{(1)}\over E}{=}
   (1{-}v_z){{\alpha_s C_R \mu^2 L^2}\over 4\lambda_gE}
   \left[\ln{2E\over \mu^2L} {-}0.048\right]{+}\mathcal{O}(|\mathbf{v}|^2)\, ,
 \end{equation}
 \begin{eqnarray}
 \label{elossab3}
 {\Delta E_{abs}^{(1)}\over E}=-
   (1-v_z){{\pi\alpha_s C_R}\over 3} {{LT^2}\over {\lambda_g
   E^2}} \left[
   \ln{{\mu^2L}\over T} -1+\gamma_{\rm E}-{{6\zeta^\prime(2)}\over\pi^2}
\right]+\mathcal{O}(|\mathbf{v}|^2)\, .
 \end{eqnarray}

Our analytic result implies, although the formation time of gluon
radiation becomes shorter, and LPM effects is reduced when $v_z>0$,
the collective flow reduces the emission current more.  To the first
order in opacity, that the energy loss is changed by a factor
$(1-v_z)$ for rescattering case with collective flow compared to the
static medium case. With collective flow velocity
$|\mathbf{v}|=0.1-0.3$ in the positive jet direction, jet energy
loss decreases by $10-30\%$. This result implies that the collective
flow has observable influence on the effective parton energy loss in
the quark-gluon medium. Our results is consistent with GLV static
potential results when the velocity of the collective flow goes to
zero. Since the parton energy loss is dominated by the first order
opacity contribution, our result also agrees with the result of
$\hat{q}$ calculation in Ref.\cite{BMS,Liu}.

\begin{figure}
\begin{center}
\includegraphics[width=80mm]{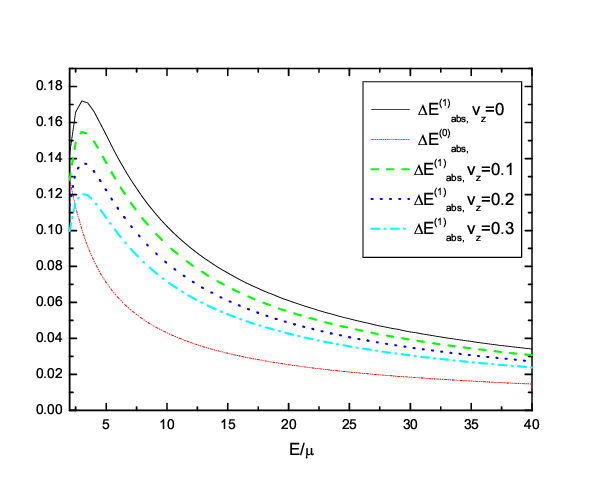}
\end{center}
\vspace{-12pt} \caption{\label{fig:ratioenergy}
  The energy gain via gluon absorption with
rescattering for $v_z=0,0.1,0.2,0.3$ and without rescattering as
functions of $E/\mu$.  }
\end{figure}

Shown in Fig. 2 are the energy gain via gluon absorption with
rescattering for $v_z=0,0.1,0.2,0.3$ and without rescattering as
functions of $E/\mu$. For comparison, we take the same values for
the medium thickness, the mean free path, and the Debye screen mass
as in Refs.\cite{GLV} and \cite{Wang:2001sf}. The energy gain
without rescattering at very small $E/\mu$ region is larger than
that with rescattering if $v_z>0.2$, but at smaller flow velocity or
at higher jet energies, it becomes smaller than that with
rescattering.

\section{Conclusion}

In summary, we have derived a new potential for the interaction of a
hard jet with the parton target. It can be used to study the jet
quenching phenomena in the presence of collective flow of the
quark-gluon medium. With this new potential, we have investigated
the effect of collective flow on jet energy loss with detailed
balance. Collective flow along the jet direction leads to a reduced
emission current square $J_{eff}^2$, $(1-v_z)^2$ times that in
static medium, and an increased LPM gluon formation factor,
$(1+v_z)$ times that in static medium. The energy gain without
rescattering is the same as in the static medium, but the total
energy loss to the first order of opacity is $(1-v_z)$ times that in
the static medium. Compared to calculations for a static medium, our
results will affect the suppression of high $p_T$ hadron spectrum
and anisotropy parameter $v_2$ in high-energy heavy-ion collisions.
Our new potential can also be used for heavy quark energy loss
calculation and will alter the dead cone effect of heavy quark jets.
Our results shall have implications for comparisons between theory
and experiment in the future.

We thank Fuqiang Wang and Xin-Nian Wang for helpful comments. This
work was supported by NSFC of China under Projects No. 10825523, No.
10635020 and No. 10875052, by MOE of China under Projects No.
IRT0624, and by MOE and SAFEA of China under Project No.
PITDU-B08033.

\end{document}